\documentclass{ws-ijmpcs}

\usepackage{ amssymb }

\def\beq{\begin{equation}}
\def\eeq{\end{equation}}

\def\<{\left \langle}
\def\>{\right \rangle}
\def\[{\left\lbrack}
\def\]{\right\rbrack}
\def\({\left(}
\def\){\right)}
\newcommand{\be}{\begin{equation}}
\newcommand{\ee}{\end{equation}}
\newcommand{\ea}{\end{eqnarray}}
\newcommand{\ba}{\begin{eqnarray}}

\newcommand{\bs}{\begin{sideways}}
\newcommand{\es}{\end{sideways}}

\newcommand{\Newt}{{\mbox{\tiny Newt}}}

\def\beq{\begin{eqnarray}}
\def\eeq{\end{eqnarray}}
\def\ln{\,\mbox{ln}\,}

\def\al{\alpha}
\def\be{\beta}

\def\de{\delta}
\def\vp{\varepsilon}

\def\ka{\kappa}

\def\na{\nabla}

\def\si{\sigma}

\def\Ga{\Gamma}
\def\De{\Delta}
\def\La{\Lambda}

\def\Om{\Omega}

\newcommand{\mnras}{Mon. Not. Roy. Astron. Soc.}
\newcommand{\apj}{Astrophys. J.}
\newcommand{\aj}{Astron. J.}

\begin{document}

\markboth{Julio C.\ Fabris,
\ \
Paulo L. C. de Oliveira
\ \
Davi C. Rodrigues,
\ \
Ilya\ L.\ Shapiro,
\ \
A.\ M.\ Velasquez-Toribio}
{(Paper's Title)}

%
\catchline{}{}{}{}{}
%

\title{ Quantum corrections to gravity
and their implications for cosmology and astrophysics}

\author{Julio C.\ Fabris,
\ \
Paulo L. C. de Oliveira
\ \
Davi C. Rodrigues,
\ \
Alan M. Velasquez-Toribio
}

\address{Departamento de F{\'\i}sica -- CCE,
\\
Universidade Federal do Esp{\'\i}rito Santo
\\
Vit\'oria, CEP: 29060-900, ES, Brazil
\\ fabris@pq.cnpq.br; \ \
davirodrigues.ufes@gmail.com; \ \
alan.toribio@ufes.br
}

\author{Ilya\ L.\ Shapiro}

\address{Departamento de F\'{\i}sica -- ICE,
\\
Universidade Federal de Juiz de Fora
\\
Juiz de Fora, CEP: 36036-330, MG,  Brazil
\\
shapiro@fisica.ufjf.br}

\maketitle

\begin{abstract}
The quantum contributions to the gravitational action are relatively
easy to calculate in the higher derivative sector of the theory. However,
the applications to the post-inflationary cosmology and astrophysics
require the corrections to the Einstein-Hilbert action and to the
cosmological constant, and those we can not derive yet in a consistent
and safe way. At the same time, if we assume that these quantum terms
are covariant and that they have relevant magnitude, their functional
form can be defined up to a single free parameter, which can be defined
on the phenomenological basis. It turns out that the quantum correction
may lead, in principle, to surprisingly strong and interesting effects
in astrophysics and cosmology\footnote{Based on the talk presented
by I. Shapiro.}.

\end{abstract}

\section{Introduction}

One of the subjects which attract a lot of attention recently is
a possible modification of General Relativity and its implications
for observational and experimental physics. Along with the long
list of possible {\it ad hoc} modifications, from the traditional
versions such as scalar-tensor theories and non-linear extensions
of the Einstein-Hilbert action, and up to the galileons and modern
versions of massive gravity, there is one which deserves, from our
viewpoint, a very special attention.
Independent on whether gravity should be or should not be quantized,
we know that the matter fields should. Therefore, it is reasonable
to ask whether the quantum effects of matter fields are capable to
produce significant effects on the astrophysical or even
cosmological scale.

At quantum level the dynamics of gravity with quantum corrections
is governed by the effective equations, coming from the Effective
Action (EA) of vacuum $\Ga[g_{\mu\nu}]$,
\beq
e^{i\Ga[g_{\mu\nu}]} = \int d\Phi
e^{iS[\Phi,\,g_{\mu\nu}]}\,,\quad
\mbox{where}
\quad
\Phi = \big\{ \mbox{matter fields} \big\}\,.
\label{EA}
\eeq

In case of renormalizable theory of matter fields we can write
\beq
S[\Phi,\,g_{\mu\nu}] = S_{vac}[g_{\mu\nu}]
+ S_m[\Phi,\,g_{\mu\nu}]
\quad
\mbox{and}
\quad
\Ga[g_{\mu\nu}]
= S_{vac}[g_{\mu\nu}] + {\bar \Ga}[g_{\mu\nu}]\,.
\label{act}
\eeq
The classical part of the vacuum action has the form
\beq
S_{vac}\,=\,S_{EH}\,+\,S_{HD}\,,\quad
\mbox{where}
\quad
S_{EH}\,=\,-\frac{1}{16\pi G} \int d^4x\sqrt{-g}\,
(R + 2\Lambda)
\label{act 1}
\eeq
and
$S_{HD}$
includes higher derivative terms.
\beq
S_{HD}\,=\,\int d^4x\sqrt{-g}\,\left\{
a_1C^2+a_2E+a_3{\Box}R+a_4R^2 \,\right\}\,.
\label{act HD}
\eeq
Here
$$
C^2(4)=R_{\mu\nu\al\be}^2 - 2R_{\al\be}^2 + 1/3\,R^2
$$
is the square of the Weyl tensor and
$$
E = R_{\mu\nu\al\be}R^{\mu\nu\al\be}
-4 \,R_{\al\be}R^{\al\be} + R^2
$$
the integrand of the Gauss-Bonnet topological invariant.

At the astrophysical or cosmological scale the quantum corrections
in the matter fields sector can not play an important role.
Therefore, from the perspective given above, the main problem is to
evaluate the quantum correction to the classical action of vacuum,
${\bar \Ga}[g_{\mu\nu}]$, at least at 1-loop.

In the case of massless conformal fields
${\bar \Ga}[g_{\mu\nu}]$
can be obtained, e.g., by integrating conformal
anomaly\cite{rie,frts84} (see also Ref.~ \refcite{Conf-Proc}
for the technical introduction and further references)
\beq
{\bar \Ga}_{ind}
&=&
S_c[g_{\mu\nu}]
+ \frac{\be_1}{4}\int_x \int_y
\left(E - \frac23{\Box}R\right)_x \,G(x,y)\,
\left(C^2\right)_y
\label{induc}
\\
&-&
\frac{\be_2}{8}\int_x \int_y \,
\left(E - \frac23{\Box}R\right)_x \,G(x,y)\,
\left(E - \frac23{\Box}R\right)_y +
\frac{3\be_3-2\be_2}{6}\int_x\,R^2\,.
\nonumber
\eeq
Here we use the notations $\int_x = \int d^4 x\sqrt{g}$
and \ $\Delta_xG(x,x^\prime)=\de(x,x^\prime)\,.$ Furthermore
\beq
\Delta = {\Box}^2 + 2R^{\mu\nu}\na_\mu\na_\nu
- \frac23\,R{\Box} + \frac13\,(\na^\mu R)\na_\mu\,,
\label{De4}
\eeq
and $S_c[g_{\mu\nu}]$ is an arbitrary conformal functional.
The $\beta$ -functions in Eq. (\ref{induc}) depend on the
number of fields, \ $N_0,\,N_{1/2},\,N_1$,
\beq
\left(
\begin{array}{c}
\be_1 \\
-\,\be_2 \\
\be_3 \end{array}\right)=\frac{1}{360(4\pi)^2}
\left(
\begin{array}{c}
3N_0 + 18N_{1/2} + 36 N_1
\\
 N_0 + 11N_{1/2} + 62 N_1
\\
2N_0 + 12N_{1/2} - 36N_1
\end{array}\right)\,.
\label{betas}
\eeq

There are many important applications of conformal anomaly
and the EA (\ref{induc}) (see, e.g., Refs.
\refcite{duff94,balsan,Reis-Nord}),
and one of the most clear ones is the Starobinsky model of
inflation\cite{star}. It is interesting that the EA
\ $S_{EH}+{\bar \Ga}_{ind}$ \ is producing
two dS-like solutions for the homogeneous and isotropic metric
(for simplicity we consider only spatially-flat case $k=0$),
\beq
ds^2 = dt^2 - a^2(t)dl^2\,,\qquad
a(t) \,=\, a_0 \cdot \exp(Ht)
\label{flat solution}
\eeq
where\cite{asta}
\beq
H\,=\, \frac{M_P}{\sqrt{-32\pi b}}\,\left(1\pm
\sqrt{1+\frac{64\pi b}{3}\frac{\Lambda }{M_P^2}}\right)^{1/2}.
\label{H}
\eeq
As far as $\,\La \ll M_P^2$, we meet two very different
values of $H$ (we consider $\La > 0$)
\beq
H_{c}\,\approx\,\sqrt{\frac{\Lambda }{3}}
\,\,\,\,\,\,\,\,\,\,\,\,
{\rm and}\,\,\,\,\,\,\,\,\,\,\,\,
H_S\,\approx\,\frac{M_P}{\sqrt{-16\pi b}}\,.
\label{HH}
\eeq
The solution with $H_c$ is the one of the theory without
quantum corrections. The second value $\,H_S\,$ corresponds
to the inflationary solution of Starobinsky\cite{star}.

Three relevant for us observations are in order.
\\
{\large $\bullet$} \ First, the expression
(\ref{induc}) is an exact EA for the conformally flat metric,
including (\ref{flat solution}) as a particular case.
\\
{\large $\bullet$} \ Second, such an exact solution is possible
only due to the very special resummation in the EA, and can not be
obtained from the massless conformal fields via the usual perturbative
approach\cite{DesSchim}. The perturbations in curvature tensor and
its covariant derivatives will always give us non-localities
related to the Green functions of the quantum fermionic, scalar
and vector fields, and not the ones of the universal
conformal operator \ $\De$ \ from (\ref{De4}).
\\
{\large $\bullet$} \ Third, nothing similar to the EA (\ref{induc})
is possible in the case of massive fields. One could think that
the effects of massive fields can not be relevant in principle at
the cosmic scale, due to the decoupling phenomenon. However, such
a statement can not be mathematically proved\cite{DCCR}. In fact,
some direct considerations using the Green functions of massive
fields indicate that the quantum effects of such fields should
be negligible, but there are two flaws in such a treatment.
First of all, it is essentially based on the expansion of the
metric near the flat background, and (as we already pointed out
here) this approach is not safe for the massive fields. For example,
the conformal parametrization - based calculations give some
positive and in fact reliable output for the case of light massive
fields\cite{Shocom,asta}. Furthermore, one can not rule out
that the EA action of massive fields can be subject of a
resummation similar to the one which leads to (\ref{EA}) in
case of anomaly-induced EA. So, after all nothing can be ruled
out completely and therefore we have a chance to meet relevant IR
vacuum quantum effects. In the rest of this contribution we shall
present a general view on the problem of vacuum quantum
effects of massive fields and also briefly discuss their possible
effects in astrophysics and cosmology. Many technical details
are omitted here, but can be found by the reader in the parallel
papers Refs.~ \refcite{RotCurves,LRL,AlFa} and
Refs.~ \refcite{paulo,ellipticals}.

\section{Covariance arguments for massive fields}

The vacuum quantum contributions of massive fields are much
more complicated and interesting, if the low-energy effects
are concerned. As we have already mentioned above, one has
to account for the decoupling phenomenon, however the result
may be different from what one could naively expect.

Let us start from the pedagogical example of QED. In the UV
limit the one-loop corrected action of photon is
\beq
-\frac{e^2}{4}F_{\mu\nu}F^{\mu\nu}
+ \frac{e^4}{3(4\pi )^2}\,F_{\mu\nu}\,
\ln \Big(-\frac{\Box}{\mu^2}\Big)\,F^{\mu\nu}
\label{1-loop}
\eeq
and we meet a standard Minimal Subtraction ($\overline{\mbox{MS}}$)
renormalization scheme based $\be$-function for $e(\mu)$. Then, at
low energies there is a quadratic decoupling, This means, in the
framework of the Renormalization Group approach, the quadratic
difference between UV and IR $\be$-function,
\beq
\mbox{UV limit}
\quad
p^2 \gg m^2 \,\,\Longrightarrow \,\,
\be_e^{1\,\,UV} \,=\,\frac{4\,e^3}{3\,(4\pi)^2}\,
+ \,{\cal O}\Big(\frac{m^2}{p^2}\Big)\,,
\qquad\quad
\\
\mbox{IR limit}\quad
p^2 \ll m^2 \,\,\Longrightarrow \,\,
\be_e^{1\,\,IR} \,=\, \frac{e^3}{(4\pi)^2}\,\cdot\,
\,\frac{4\,p^2}{15\,m^2} \,\,
+ \,\,{\cal O}\Big(\frac{p^4}{m^4}\Big)\,,
\eeq
that is the Appelquist and Carazzone decoupling theorem\cite{AC}.

Similar results can be obtained for gravity, e.g., for a massive
scalar field we have the following UV and IR $\be$-functions for
the parameter $a_1$ in the action (\ref{act HD})\cite{apco,fervi}:
\beq
\be_1^{UV} &=& -\frac{1}{(4\pi)^2}\frac{1}{120}
+ {\cal O}\left(\frac{m^2}{p^2}\right)
= \be_1^{\overline {MS}}
+ {\cal O}\left(\frac{m^2}{p^2}\right),
\nonumber
\\
\be_1^{IR} &=& -\,\frac{1}{1680\,(4\pi)^2}\,\cdot\,
\frac{p^2}{m^2}
\,+\,{\cal O}\left(\frac{p^4}{m^4}\right)\,,
\label{C2}
\eeq
This is the Appelquist and Carazzone Theorem for gravity,
it implies a quadratic suppression of the running in the IR.
The same rule holds also for spin-$1/2$ and spin-$1$ fields for
both $C^2$ and $R^2$ terms. All these results were obtained
through the momentum-subtraction scheme, in the flat-space
expansion $g_{\mu\nu}=\eta_{\mu\nu}+h_{\mu\nu}$ or in an
equivalent perturbative (in curvatures) heat-kernel approach.
However, it is easy
to see that in the momentum-subtraction scheme \
$\be_{1/G} = \be_\La = 0$, because the $\Box$-dependent
form factors like the one of Eq. (\ref{1-loop}),
can not be inserted into the Hilbert-Einstein
and cosmological terms\cite{apco}.  \
At the same time there is no problem to insert such a
form factor into $C^2$ term,
\beq
C_{\mu\nu\al\be}\,
\ln \Big(-\frac{\Box}{\mu^2}\Big)\,C_{\mu\nu\al\be}\,.
\nonumber
\eeq
and similarly to $R^2$ term, that is why we can
study the running of the corresponding parameters in the
momentum-subtraction scheme of renormalization.

From the consideration presented above, it becomes clear
why we get an apparent $\beta_{\Lambda}=\beta_{1/G}=0$ in
the momentum-subtraction scheme. The reason is that the
expansion $g_{\mu\nu}=\eta_{\mu\nu}+h_{\mu\nu}$ is not
appropriate for the massive fields case. The renormalizable
theory of massive fields {\it has to} include the cosmological
constant term in this case, and then $\eta_{\mu\nu}$ is not a
classical solution anymore. In this situation the expansion
around flat space is not a legitimate procedure.
Perhaps the linearized gravity approach is simply
not an appropriate tool for the CC and Einstein terms.
If we perform some other expansion, the output for the
$\beta_{\Lambda}$ and $\beta_{1/G}$ can be different, but
this is out of our knowledge at the moment.

As far as the direct theoretical derivation of the quantum
effects of our interest is not possible, we can look at the
problem from the phenomenological side. One can simply make
an assumption that some relevant quantum contributions are
present, and then use the covariance arguments to find their
form. Later on we will see how this approach fits also to
the Appelquist and Carazzone theorem.

Consider first the cosmological term and perform a derivative
expansion in the EA. The EA $\Ga[g_{\mu\nu}]$ can not include
odd terms in metric derivatives, just because it is a covariant
scalar. In the cosmological setting this means there are no
${\cal O}(H)$ terms, and also no ${\cal O}(H^3)$ and so
on\cite{PoImpo}. Hence the covariance arguments give the
formula
\beq
\rho_\La(H) &=& \frac{\La(H)}{16\pi G(H)}
\,=\, \rho_\La(H_0)
\,+\, C\,\big( H^2-H^2_0\Big)\,,
\label{RG-CC-cosm}
\\
\mbox{where} &&
C = \frac{3\nu}{8\pi}\,M_p^2\,\big( H^2-H^2_0\Big)
\nonumber
\eeq
and the physical sense of the constant parameter $\nu$
will de defined later on, in Eq. (\ref{RG CC}). Starting from
(\ref{RG-CC-cosm}) the standard covariance (conservation law)
consideration leads to the relation\cite{Gruni}
\beq
G(H;\nu)=\frac{G_0}{1+\nu\,\ln\left(H^2/H_0^2\right)}\,,
\quad \mbox{where}  \quad
G(H_0) = G_0 = \frac{1}{M_P^2}\,.
\label{RG-G-cosm}
\eeq
From the renormalization group perspective, the identification
of scale $\mu \,\sim\,H$ is the most natural in the cosmological
setting\cite{babic,nova}\footnote{See Ref. \refcite{fossil} for
an alternative treatment.}. Therefore the last formulas can be
generalized as
\beq
\rho_\La(\mu) \,=\, \rho_\La(\mu_0)
\,+\, C\,\big( \mu^2-\mu^2_0\Big)\,.
\qquad
G(\mu)=\frac{G_0}{1+\nu\,\ln\left(\mu^2/\mu_0^2\right)}\,,
\label{RG}
\eeq
where $\mu_0$ is the reference scale. We will discuss an
identification of $\mu$ for the astrophysical case below.

Before we proceed, it is worthwhile to make a small note on
the Cosmological Constant (CC) Problem\cite{weinberg89,nova}.
The main relation, from the QFT viewpoint, is that the
observed density of the cosmological constant term is a
sum of the two finite terms, namely of the vacuum and
induced one,
\beq
\rho_\La^{obs}\,=\,\rho_\La^{vac}(\mu_c) + \rho_\La^{ind}(\mu_c)\,.
\label{CC}
\eeq
where $\mu_c \propto H_0$ is the late Universe cosmic scale. Here $\rho_\La^{obs}$ is the value which is likely observed in SN-Ia,
LSS, CMB etc, to be \ \ $\rho_\La^{obs}(\mu_c) \approx
0.7\,\rho_c^0 \,\propto \, 10^{-47}\,GeV^4$. \ The unusual
feature of the relation (\ref{CC}) is that the two terms
$\,\rho_\La^{vac}(\mu_c)\,$
and
$\,\rho_\La^{ind}(\mu_c)\,$
in the {\it r.h.s.} are evaluated at the (at least) Fermi
scale and therefore have much greater magnitudes, of at least
\ $10^{8}\,GeV^4$. The main CC Problem is that these magnitudes
of are a huge 55 orders of magnitude greater than the sum.
Obviously, these two huge terms do cancel. Here we follow a
phenomenological attitude and don't try solving the main CC
problem. Instead we consider whether CC may vary due to IR
quantum effects of massive matter fields.

It is remarkable that the same equation (\ref{RG})
follows from the assumption of the Appelquist and
Carazzone - like decoupling for CC\cite{nova}. For a single
particle the $\be$-function for $\,\rho_\La^{vac}(\mu)\,$ is
$$
\be^{MS}_\Lambda (m)\,\sim\, m^4\,,
$$
hence the quadratic decoupling gives
$$
\be^{IR}_\Lambda (m)\,=\,
\frac{\mu^2}{m^2}\,\be^{MS}_\Lambda (m)\,\sim\, \mu^2 m^2\,.
$$
Then the total beta-function will be given by an algebraic sum
$$
\be^{IR}_\Lambda \,=\,
\sum\, k_i\mu^2 m_i^2  \,=\, \si M^2 \,\mu^2
\,\propto\, \frac{3\nu}{8\pi}\, M_P^2 \,H^2 \,.
$$
This leads to the same result (\ref{RG-CC-cosm}), in the cosmological
setting,
\beq
\rho_\La(H) \,=\, \rho_\La(H_0)
\,+\, \frac{3\nu}{8\pi}\,M_p^2\,\big( H^2-H^2_0\Big)\,.
\label{RG CC}
\eeq

It is also remarkable that one can also obtain the same $G(\mu)$,
Eq. (\ref{RG}), in one more independent way\cite{nova,LRL}. Consider
$\overline{\rm MS}$-based renormalization group equation for \ $G(\mu)$,
\beq
\mu \frac{dG^{-1}}{d\mu}\,=\,\sum\limits_{particles}
\,A_{ij}\,m_i\,m_j\,=\,2\nu\,M_P^2 \,,\qquad
G^{-1}(\mu_0)=G^{-1}_0=M^2_P\,.
\label{hip 1}
\eeq
Here the coefficients $A_{ij}$
depend on the coupling constants, \ $m_{i}$ \
are masses of all particles. In particular, at one loop,
$$
\sum\limits_{particles}A_{ij}\,m_i\,m_j
\,=\,\,\sum\limits_{fermions} \frac{m_f^2}{3(4\pi)^2}
\,-\,\sum\limits_{scalars} \frac{m_s^2}{(4\pi)^2}
\Big(\xi_s-\frac16\Big)\,.
$$
One can rewrite Eq. (\ref{hip 1}) as
\beq
\mu \frac{d(G/G_0)}{d\mu}\,=\,-\,2\nu\,(G/G_0)^2
\quad\Longrightarrow\quad
G(\mu)=\frac{G_0}{1 + \nu\,\ln\left(\mu^2/\mu_0^2\right)}\,.
\label{hip 2}
\eeq
It is easy to see that we arrived at the same formula (\ref{RG}),
which results from covariance arguments and/or from Appelquist
and Carazzone-like
quadratic decoupling for the CC plus conservation law.
Eq. (\ref{hip 2}) is the unique possible form of a relevant
running $G(\mu)$. An alternative to this relation is the
non-running, that means simply $\nu=0$.

From the perspective described above, it is not a surprise that
the Eq. (\ref{hip 2}) emerges in very different approaches to
renormalization group in gravity, including higher derivative
quantum gravity\cite{Salam,frts82}; non-perturbative quantum
gravity with (hypothetic) UV-stable fixed point\cite{bonreuter}
and semiclassical gravity\cite{nelspan82,book}.

As far as we arrived at the two relations (\ref{RG-CC-cosm}) and
(\ref{RG-G-cosm}) in the cosmological setting, it is natural to
construct cosmological models based on these formulas. The first
steps in this directions has been done in Ref.~ \refcite{CCfit}
where the cosmological models with energy matter-vacuum exchange
and constant $G$ were constructed in Ref.~ \refcite{Gruni}, where
the cosmological model without matter-vacuum exchange was
constructed by assuming the scale-dependence running
(\ref{RG-G-cosm})
for $G$. In this presentation we will not describe the details
of these models. Let us only mention that the density perturbations
were explored for these models by different
methods\cite{OphPel,WangMeng,CCwave,CCG}. In particular, the
result of Ref.~ \refcite{CCG} implies that the possible quantum
contributions (\ref{RG}) do not really affect the power
spectrum of the cosmological model, such that the last remains
almost the same as in the classical case. We will discuss the
importance of this result in Sect. 4.

\section{Galaxies}

If the quantum effects parametrized by Eqs. (\ref{RG}) really
take place and are relevant even at the scale of the whole
universe, they can manifest themselves also at the astrophysical
scale. What could be an interpretation of $\mu$ in astrophysics?

Consider the rotation curves of galaxies. The simplest assumption
is $\mu \propto 1/r$, and this identification has been applied
for the point-like model of galaxy in
Refs.~ \refcite{bertol,reuter-04}
and Ref.~ \refcite{Gruni}. In fact, the method suggested in
Ref.~ \refcite{Gruni} (see also Ref.~ \refcite{RotCurves}) is
quite general, and can be used for various identifications of $\mu$.
The main idea is to consider a weakly varying
\beq
G = G_0 + \de G = G_0(1+\ka)\,\qquad
\left|\ka \right| \ll 1
\label{G weak}
\eeq
and perform a conformal transformation
\beq
{\bar g}_{\mu \nu} = \frac{G_0}{G} ~ g_{\mu \nu}
= (1-\ka)g_{\mu \nu}\,.
\label{G con}
\eeq
It is easy to see that in the first order in $\ka$ the metric
${\bar g}_{\mu \nu}$ satisfies usual Einstein equations with
constant $G_0$.

The nonrelativistic limits of the two metrics are
\beq
g_{00} = - 1 -  \frac {2 \Phi}{c^2}
\quad \mbox{and} \quad
{\bar g}_{00}=-1- \frac{2 \Phi_\Newt}{c^2} \quad\,,
\label{two}
\eeq
where $\Phi_\Newt$ is the usual Newton potential and
$\Phi$ is a potential of the modifies gravitational theory.

For the nonrelativistic limit of the modified gravitational
force we obtain, in this way,
\beq
-\Phi^{,i}
\,=\, - \Phi^{,i}_{\Newt} \,-\, \frac{c^2 \, G^{,i}}{2\,G_0}\,.
\label{twotwo}
\eeq

The formula (\ref{twotwo}) is very instructive.
Quantum correction comes multiplied by \ $c^2$ \ and therefore
it does not need to be very big to make real effect even
at the galaxy scale. For a point-like model of galaxy and
$\mu \propto 1/r$ it is sufficient to have $\nu \approx 10^{-6}$
to provide the flat rotation curves\cite{Gruni}.
At the same time ref.  (\ref{twotwo}) shows that it not a
really good choice for a non-point-like model of the galaxy.
The reason is that this identification produces the
``quantum-gravitational'' force even if there is no mass
at all.

What would be the ``right''
identification of the renormalization group scale parameter
in the almost-Newtonian regime? Let us come back to the (QFT).
Then it is clear that $\mu$ must be associated to some
parameter which characterizes the energy of the particle
which is transmitting gravitational interaction.
Of course, $\mu \propto 1/r$ is not the right choice.

The phenomenologically good choice is
\beq
\frac {\mu}{\mu_0}
=  \Big( \frac{\Phi_{\mbox{\tiny Newt}}}{\Phi_0} \Big)^\alpha\,,
\label{three}
\eeq
where  $\alpha$ \ is a phenomenological parameter which can
be distinct for different spiral galaxies. We have found that
\ $\alpha$ \ is nonlinearly growing with the mass of
the galaxy.

From the QFT viewpoint the presence of $\al$ reflects the
fact that the association of $\mu$ with
$\Phi_{\mbox{\tiny Newt}}$ is not an ultimate choice.
Remember that the vacuum EA is a relativistic object and
taking $\Phi_{\mbox{\tiny Newt}}$ as a scale definitely
ignores some relevant information.
With greater mass of the galaxy the
``error'' in identification becomes greater too, hence
we need a greater $\al$ to correct this.
Furthermore, if $\al$ increase with the mass of the galaxy,
it must be very small at the scale of the Sun system
and of course at the scale of laboratory, when the Newton
law is better verified. Finally, the recently-proposed
regular scale-setting procedure gives the very same
result\cite{DomStef}.

In Ref.~ \refcite{RotCurves} we applied the RGGR model to nine
disk galaxies (including high and low surface brightness galaxies)
from two sample of data\cite{things}\cdash\cite{gentile}. We have
also compared our results to three other models: a model with a
dark matter halo given by the phenomenologically successful
(pseudo-)isothermal profile; the Modified Newtonian Dynamics
(MOND)\cite{milgrom} (in its original form); and the Metric Skew
Tensor Gravity (MSTG)\cite{BrownsteinMoffat}. For the shape of
the rotation curve, the RGGR model has in general achieved lower
$\chi^2$ and $\chi^2_{\mbox{\tiny reduced}}$ than MOND and
MSTG\footnote{The isothermal profile has one more free parameter
than RGGR, while the latter has one more free parameter than MOND
and MSTG.}.
Considering the expected mass to light ratio, as inferred from the
Kroupa initial mass function (IMF) as derived in Ref.~ \refcite{things},
the RGGR model has achieved better results than all the other proposals.
\\
\begin{figure}[pb]
\centerline{\psfig{file=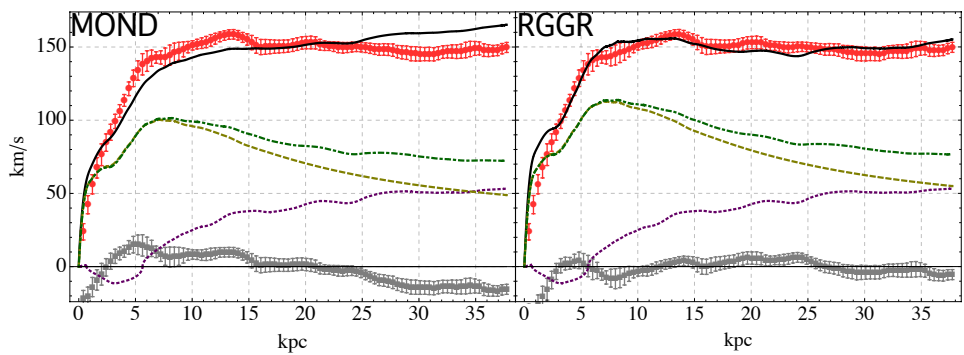,width=12.7cm}}
\vspace*{8pt}
\caption{\small NGC 3198 rotation curve fits. On the left is
the fit with the original MOND prescription, and on the right
the RGGR model. The upper dots and its error bars are the
rotation curve observational data. The lower dots with error
bars are the residues of the fit. The solid black line for
each model is its best fit rotation curve, the dashed  curves
are the stellar rotation curves,  the dotted  curve is the gas
rotation curve, and the dot-dashed  curve is the resulting
Newtonian rotation curve (both without dark matter). }
\label{n3198}
\end{figure}
\\
Since the results were very good, we are both extending our sample
of disk galaxies\cite{paulo} and applying RGGR to elliptical
galaxies\cite{ellipticals}. This is important both to show that
RGGR can indeed work for a larger sample and to unveil, in particular,
the behavior of the $\alpha$ parameter from system to system.
\\
\begin{figure}[pb]
\centerline{\psfig{file=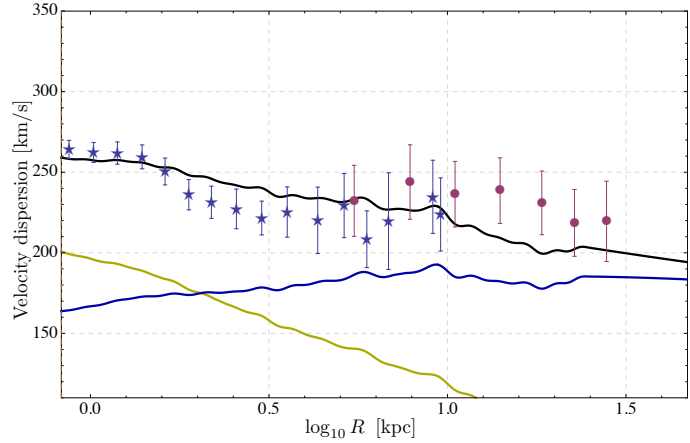,width=9.0cm}}
\vspace*{8pt}
\caption{\small RGGR mass modeling of the giant elliptical galaxy
NGC 4374. The stars with the error bars are the velocity dispersion
(VD)  data found from the stellar photometry. The filled circles
with error bars are the velocity dispersion data found from
more than 450 planetary nebulae. The upper curve is the resulting
best fit VD curve for the RGGR model. The descending line at the
bottom is the stellar VD curve, while the crescent curve is the
additional (non-Newtonian) RGGR contribution to the dispersion.
The square of the last two is equal to the square of the resulting
(upper) curve.  }
\label{n4374}
\end{figure}
\\
Besides testing RGGR in a larger sample of disk galaxies, we are
also testing the robustness of the model results once different
assumptions on the baryonic matter are done. In particular, while
at Ref.~ \refcite{RotCurves} we used exponential approximations to
the matter distribution of all galaxies at all radii, we are
modeling again some of the previous galaxies but using the
photometric data up to the radius such data is known. An example
of the newer results can be seen in Fig. \ref{n3198}. Comparing
this result with the corresponding one in Ref.~ \refcite{RotCurves},
it can be seen that there is no considerable difference in both the
shape and the inferred parameters for this case (apart from the
shape of the central region, which poses difficulties to any
model\cite{things}). For this galaxy (whose data come from
Ref.~ \refcite{things}), the inferred mass-to-light ratio and the
value of $\alpha \nu$ are essentially the same of those found
in Ref.~ \refcite{RotCurves} ($Y^{\mbox{\tiny 3.6}}_* \approx 0.8$
and $\alpha \nu  \approx 1.7\times 10^{-7}$, with
$\chi_{\mbox{\tiny reduced}} = 1.9$).
In the same figure, from exactly the same data, we also show the
result of applying MOND in its original form
($Y^{\mbox{\tiny 3.6}}_* \approx 0.7 \frac{L_\odot}{M_\odot},
\chi_{\mbox{\tiny reduced}} = 5.5$). The discrepancy found from
MOND directly applied to the NGC 3198 current data is a well
known issue, see in particular the recent comments in
Ref.~ \refcite{thingsmond}.
\\
\begin{figure}[pb]
\centerline{\psfig{file=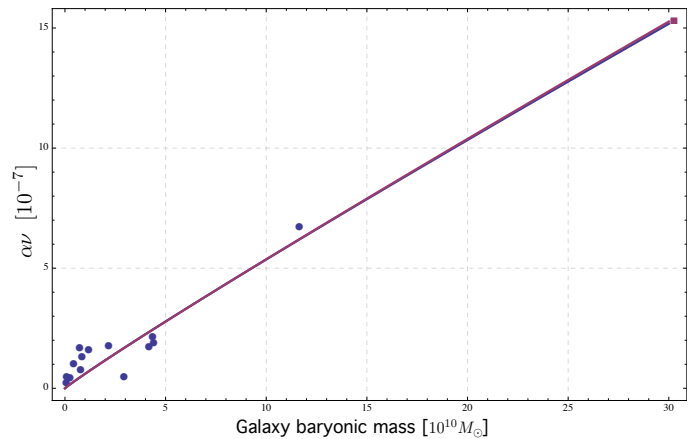,width=9.0cm}}
\vspace*{8pt}
\caption{\small A preliminary plot showing how the parameter $\alpha$
varies from galaxy to galaxy, considering their baryonic (stellar and
gaseous) masses alone. Since galaxy dynamics is only sensible to
$\alpha$ through the product $\alpha \nu$, in the above the total
baryonic mass of  15 galaxies are plotted against $\alpha \nu$; 14
of them are disk galaxies (depicted by circles), and one is a giant
elliptical galaxy (NGC 4374), depicted by a square. There are two
lines in the above plot that are almost identical. Both were derived
from the best fit of $\alpha = a \; \mbox{Mass}^b$, where $a$ and $b$
are constants, but one of them considering only the disk galaxies and
the other all the 15 galaxies. For this range of masses, $b \approx 1$
is a solution of the fit, regardless whether  the elliptical NGC 4374
is part of the data or not.}
\label{allgalaxies}
\end{figure}
\\
Figure \ref{n4374} shows the application of RGGR to the mass
modeling of the giant elliptical galaxy NGC 4374, using the same
data of Ref.~ \refcite{napolitano2011}. To derive this dispersion
curve, it is only necessary to replace the effective potential
in the Jeans equation by the RGGR potential derived from the stellar
density\cite{ellipticals}. The resulting fit is very good: assuming
constant anisotropy (constant $\beta$  in the Jeans equations), the
RGGR model automatically leads to $\beta \approx 0.1 $ (i.e., close
to isotropic), to a mass-to-light ratio in perfect agreement with
the Kroupa IMF\cite{napolitano2011},
$Y_*^V \approx 4 \frac{L_\odot}{M_\odot}$,
$\nu \alpha \approx 15 \times 10^{-7}$
(in agreement with its expectation of becoming larger for larger
masses) and $\chi_{\mbox{\tiny reduced}} = 1.0$.

In order to start to disclose the parameter $\alpha$ relation to
the system mass, we show our partial results in the Fig.
\ref{allgalaxies}. The result shown in this plot is consistent
with the bound $|\alpha \nu| < 10^{-17}$ for the solar system,
which was recently derived for the Solar system in
Ref.~\refcite{LRL} by using the weak non-conservation of the
Laplace-Runge-Lenz vector. Further details concerning the
analysis of all worked out galaxies will be available soon in
Ref.~ \refcite{paulo}.

\section{Cosmological applications}

It looks like we do not need CDs to explain the rotation curves
of the galaxies. However, does it really mean that we can
really go on with one less dark component? Maybe at the end
of the day the answer will be negative, but it is definitely
worthwhile to check such a possibility.

It is well known that the main
requests for the DM come from the fitting of the LSS, CMB,
BAO, lensing etc. However there is certain hope to replace,
e.g., $\La$CDM by a WDM, e.g., by sterile neutrino with much
smaller $\Omega_{DM}$. So, the idea it to trade the set of
\ $(\Om_{BM},\,\Om_{DM},\,\Om_{CC})$ \ from the conventional
\ $(0.04,\,0.23,\,0.73)$ \ to \ $(0.04,\,\,0.0x,\,\,0.9(1-x))$
with
a relatively small $x$. Such a new concordance model would
have less relevant coincidence problem, and in general
such a possibility is interesting to verify. The first move
in this direction has been done recently in Ref.~ \refcite{AlFa}
by using the Reduced Relativistic Gas (RRG) model.

The RRG model is a Simple cosmological model of a universe
filled by ideal relativistic gas of massive particles\cite{FlaFlu}.
As an approximation we assume that all of these particles
have the same kinetic energy. The Equation of State (EOS) of
such gas is\cite{sakh,FlaFlu}
\beq
P = \frac{\rho}{3}\,\Big[1-\Big(\frac{mc^{2}}{\vp}\Big)\Big]^{2}
\,=\, \frac{\rho}{3}\,\Big(1-\frac{\rho^2_{d}}{\rho^2}\Big)\,.
\label{RRG-EOS}
\eeq
In this formula $\vp$ is the kinetic energy of the individual
particle, $\vp=mc^2/\sqrt{1-\be^2}$. Furthermore,
$\rho_{d}=\rho_{d0}^{2}(1+z)^{3}$ is the mass (static energy)
density. One can use one or another form of the
equation of state (\ref{RRG-EOS}), depending on the situation.
The nice thing is that one can solve the
Friedmann equation in this model analytically. The deviation
from Maxwell or relativistic Fermi-Dirac distribution is less
than 2.5\%.
It is amusing that the same EOS has been used in
Ref.~ \refcite{sakh} by A.D. Sakharov in 1965 to predict
the oscillations in the CMB spectrum for the first time.

In Ref.~ \refcite{AlFa} we have used RRG without quantum effects to
fit such sets of observational data as Supernova type Ia
(Union2 sample), $H(z)$, CMB ($R$ factor), BAO and LSS
(2dfGRS). Taking all these tests together we confirm that
the $\La$CDM is definitely the most favored model. As far
as we tried the model {\it without} quantum effects, this
output can be seen as a successful test of RRG and nothing
else.

However, there is a very important extra detail which
concerns the LSS part alone. In this case we met the
possibility of an alternative model with a small quantity
of a WDM. This output is potentially relevant in view of the
fact that (as we have already emphasized above) the LSS is
the only test which can not be affected by the possible
quantum renormalization-group running in the low-energy
gravitational action. Let us present here a few details
of the results of Ref.~ \refcite{AlFa}.

\begin{figure}[!h]
\begin{center}
\includegraphics[height= 7. cm,width=6.0cm]{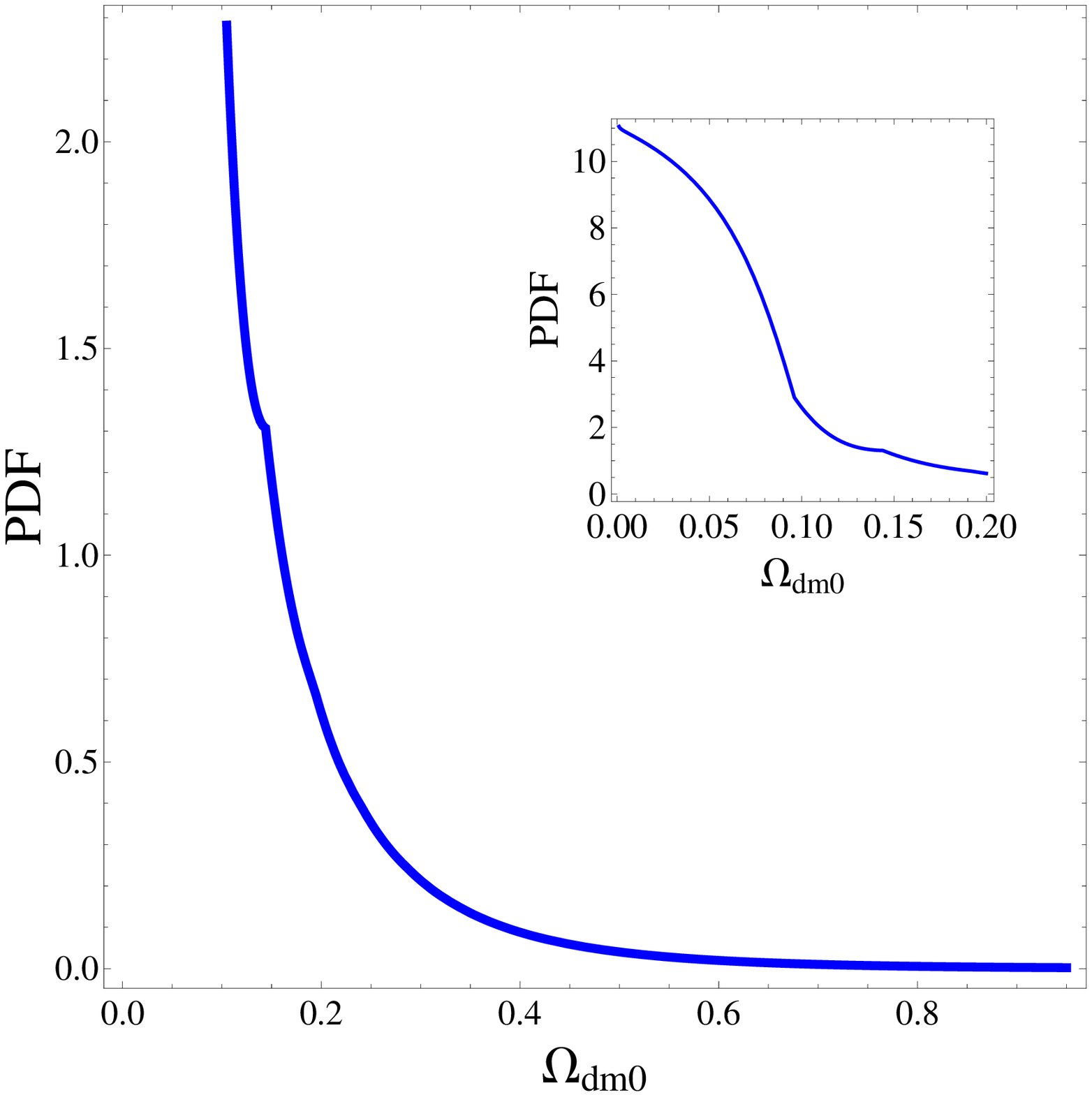}
\includegraphics[height= 7. cm,width=6.5cm]{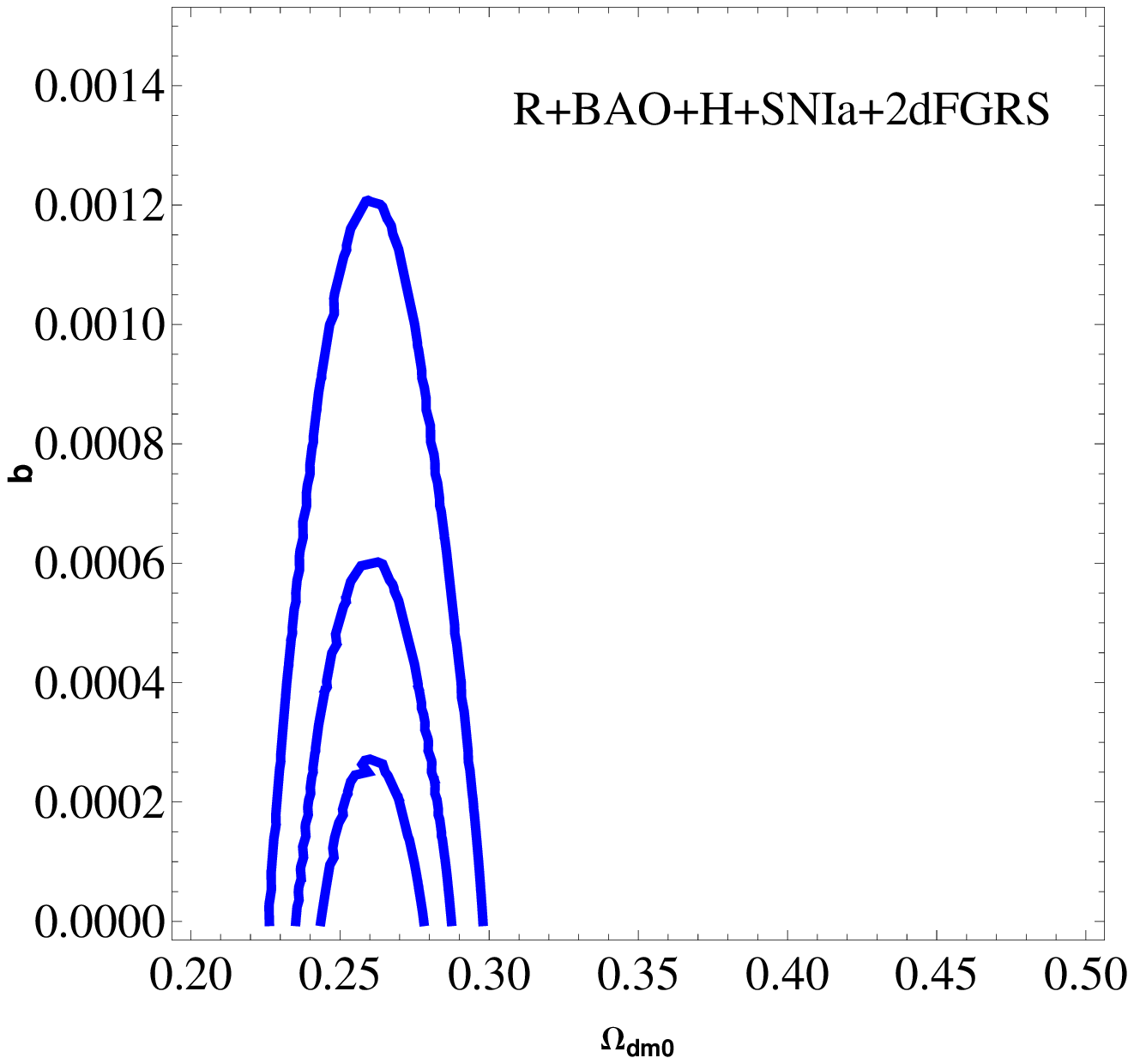}
\end{center}
\caption{The probability density function using 2dFGRS for
$\Omega_{m0}$(left).
To build the PDF we marginalize the
free parameters considering the intervals:
$\Omega_{\rm dm0}$ $\epsilon$ $[0.05, 0.95]$
 and $b$ $\epsilon$
$[0.001, 0.4]$,
for each case. As we can see the  model is included,i.e,
the first figure shows that for a $\Omega_{\rm dm0} \approx$ $0.25$
corresponds to a $b \approx 0$.
Confidence regions at 1-$\sigma$, 2-$\sigma$ and 3-$\sigma$
levels(right) from
inner to outer respectively on the ($\Omega_{\rm dm0}$ , $b$) plane for
our relativistic model in the flat case.}
\label{fig1-15}
\end{figure}

The cosmological model based on RRG with the presence of the
cosmological constant admits an analytic solution for the
energy density. This solution does interpolates between
the radiation-dominated and the matter-dominated
eras\cite{sakh,FlaFlu}.
It can represent a warm dark matter(WDM), characterized by the
parameter, $b=\frac{\beta}{\sqrt{1+\beta^{2}}}$.

In Ref.~ \refcite{sWIMPs} the model was successfully used to make a
analysis of density perturbations and comparison with the 2dFGRS
data. Using the RRG model to derive and analyze density
perturbations at the linear level one arrives at the conclusion
that the upper bound for the warmness parameter is
$b \leq 3-4 \times 10^{-5}$.
It is about two order of magnitude smaller than the escape
velocities  for the spiral galaxies. This result was similar to
those obtained using non-analytical models of WDM based on the
system of Boltzmann-Einstein equations.

As we have already mentioned, in Ref.~ \refcite{AlFa}
the model has been tested using four Supernova type Ia (Union2 sample),
$H(z)$, CMB ($R$ factor) and BAO. Moreover, a detailed study of
structure formation at linear level has been performed using the
2dFGRS data for matter power spectrum. The different tests have
been crossed in order to obtain a more clear evaluation for the
free parameters, which are essentially the velocity
parameter for the dark matter particles $b$ and the dark matter
ratio to the critical density $\Omega_{\rm dm0}$. All the analysis
has been performed using the flat universe prior. In general,
we confirm that $\Lambda$CDM is the most favored model. However,
for the LSS data the maximum probability for $\Omega_{\rm dm0}$
occurs at a zero value. This seems to be a consequence of the
restriction of the analysis to a linear level, since a certain
amount of dark matter is necessary in order to have the formation
of structure process. In any case, a small amount of dark matter
is certainly admitted much less than that predicted by the
$\Lambda$CDM model. The results are shown in Fig. 1.
Therefore, for the 2dFGRS data alone we met the possibility of an
alternative model with a small quantity of a WDM.
This output is potentially relevant in view of the fact that the LSS
is the only test which can not be affected by the possible
quantum renormalization-group running in the low-energy
gravitational action.

\section{Conclusions and discussions}

The evaluation of quantum corrections from massive fields is,
to some extent, reduced to existing-nonexisting paradigm. There
is no theoretical way to prove or disprove the existence of such
quantum corrections\cite{DCCR} and on has to rely on faith or
use phenomenological approach, that means simply assume the
existence of such quantum corrections and check their possible
consequences. In this way we arrive at the cosmological and
astrophysical model with one free parameter plus certain freedom
of scale identification. It turns out that the rotation curves
of all tested galaxies can be described by the $G(\mu)$ formula.
The situation with clusters and other tests, especially CMB and
gravitational lensing, remains unclear, because it was not
explored at all. At the same time, we have a very strong positive
signal from the analysis of the LSS data. The power spectrum
tests are almost not sensible to the $G(\mu)$ running and,
exactly in this case, we meet an alternative to $\La$CDM
in the zero-order approximation\cite{AlFa}.

Finally, we can conclude that there is still some (albeit
they can be evaluated to be small) chance that the vacuum
effects of QFT in an external gravitational field play more
significant role in our Universe that we usually think.
In particular, we gain a chance to resolve the so-called
coincidence problem for the CC in a qualitatively new way.
This problem consists in the question of why our Universe
is such that the cosmic acceleration has started only recently.
However, if the present-day $\Om_{CC}$ is more than $0.9$,
the moment when this acceleration starts move essentially
back to the past and there is no such question. Of course,
many tests of the possible cosmic-scale effects of quantum
corrections are necessary before one can think about this
solution seriously, but the results of preliminary studies
described here indicate that the subject is interesting and
it is worthwhile to study it in more details, from both
phenomenological and theoretical sides.

\section*{Acknowledgments.}
I.Sh. is very grateful to the Organizers of the QFEXT-2011
meeting in Benasque for invitation and for the possibility
to present our work as a plenary talk.
We thank W.J.G. de Blok and N.R. Napolitano for providing us
observational data on the galaxies NGC 3198 and NGC 4374.
The work of I.Sh. has been partially supported by CNPq,
FAPEMIG and ICTP. J.F. was partially supported by CNPq.


\end{document}